\documentclass[onecolumn,showpacs,preprintnumbers,amsmath,amssymb,floatfix]{revtex4}

\usepackage{graphicx}% Include figure files
\usepackage{dcolumn}% Align table columns on decimal point
\usepackage{bm}% bold math

\newcommand{\pom}{\tt I\! P}
\newcommand{\beq}{\begin{equation}}
\newcommand{\eeq}{\end{equation}}

\def \pom {{I\!\!P}}

\begin{document}
\begin{flushright}
LU TP 15-54\\
November 2015
\vskip1cm
\end{flushright}

\title{Bottom production in Photon and Pomeron -- induced interactions at the LHC}
\author{V. P. Goncalves$^{1,2}$, C. Potterat$^{3}$, M. S. Rangel$^{3}$ }
\affiliation{ $^{1}$ Department of Astronomy and Theoretical Physics, Lund University, 223-62 Lund, Sweden.}
\affiliation{$^{2}$ Instituto de F\'{\i}sica e Matem\'atica,  Universidade
Federal de Pelotas, %\\
Caixa Postal 354, CEP 96010-900, Pelotas, RS, Brazil}
\affiliation{$^{3}$ Instituto de F\'isica, Universidade Federal do Rio de Janeiro (UFRJ), 
Caixa Postal 68528, CEP 21941-972, Rio de Janeiro, Brazil\\}

\begin{abstract}
In this paper we present a detailed comparison of  the bottom  production in gluon -- gluon, photon -- gluon, photon -- photon, pomeron -- gluon, pomeron -- pomeron and pomeron -- photon interactions at the LHC. The transverse momentum, pseudo -- rapidity and $\xi$ dependencies of the cross sections are calculated at LHC energy using the Forward Physics Monte Carlo (FPMC), which allows to obtain realistic predictions for the bottom production with one or two leading intact protons. Moreover, predictions for the the kinematical range probed by the LHCb Collaboration are also presented. Our results indicate that the analysis of the single diffractive events is feasible using the Run I LHCb data.

\end{abstract}


\maketitle

\section{Introduction}

Heavy quark production in hard collisions  is considered as a clean test of perturbative Quantum Chromodynamics (QCD)  (For a review see, e.g., Ref \cite{review_hq}). This process provides not only many tests of perturbative QCD, but also some of the most important backgrounds to new physics processes, which have  motivated the development  of an extensive phenomenology at DESY-HERA,  Tevatron and  LHC. Furthermore, heavy quark production also is considered a  good testing ground for diffractive physics and for the nature of the Pomeron ($\pom$), which is a long-standing puzzle in the particle physics \cite{pomeron}. In particular, if the Pomeron is assumed to have a partonic structure, as proposed 
by Ingelman and Schlein  \cite{IS} many years ago, the production of heavy quarks in diffractive processes, is a direct probe of the   quark and gluon content in the Pomeron \cite{Engel,MMM1,antoni,antoni2,victor}.

\begin{figure}[t]
\begin{center}
\begin{tabular}{ccc}
\scalebox{0.18}{\includegraphics{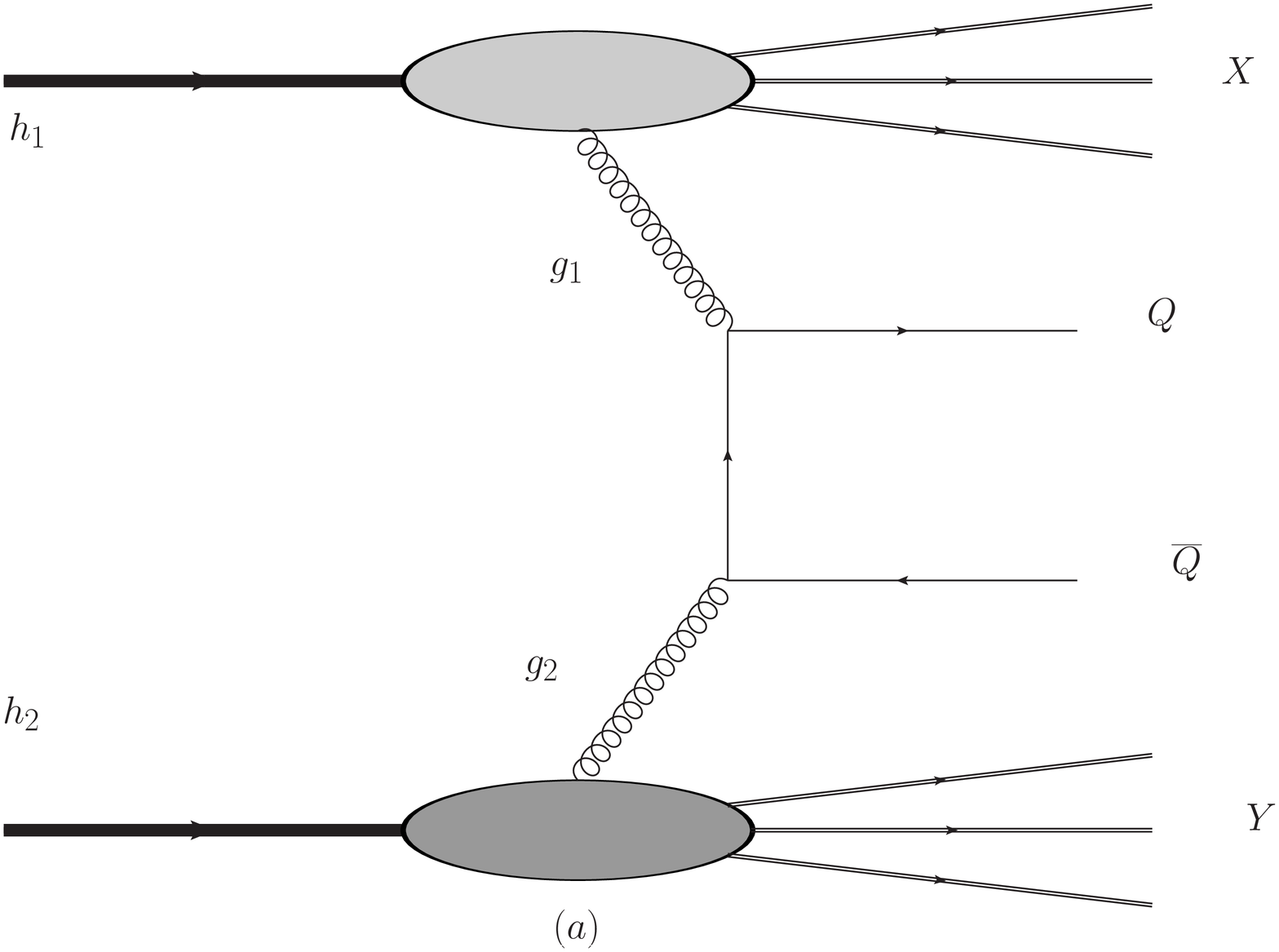}}&\scalebox{0.18}{\includegraphics{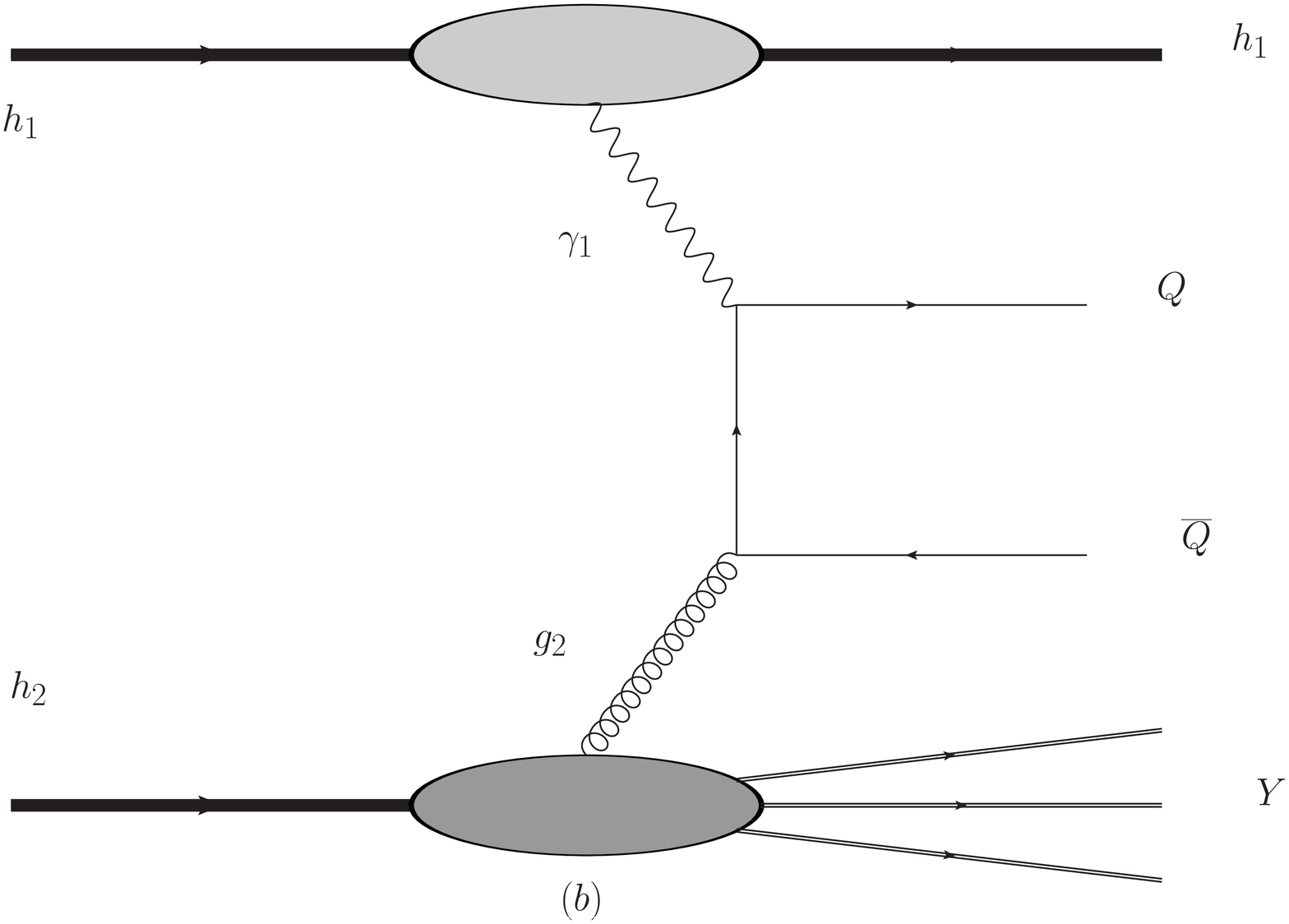}} &\scalebox{0.18}{\includegraphics{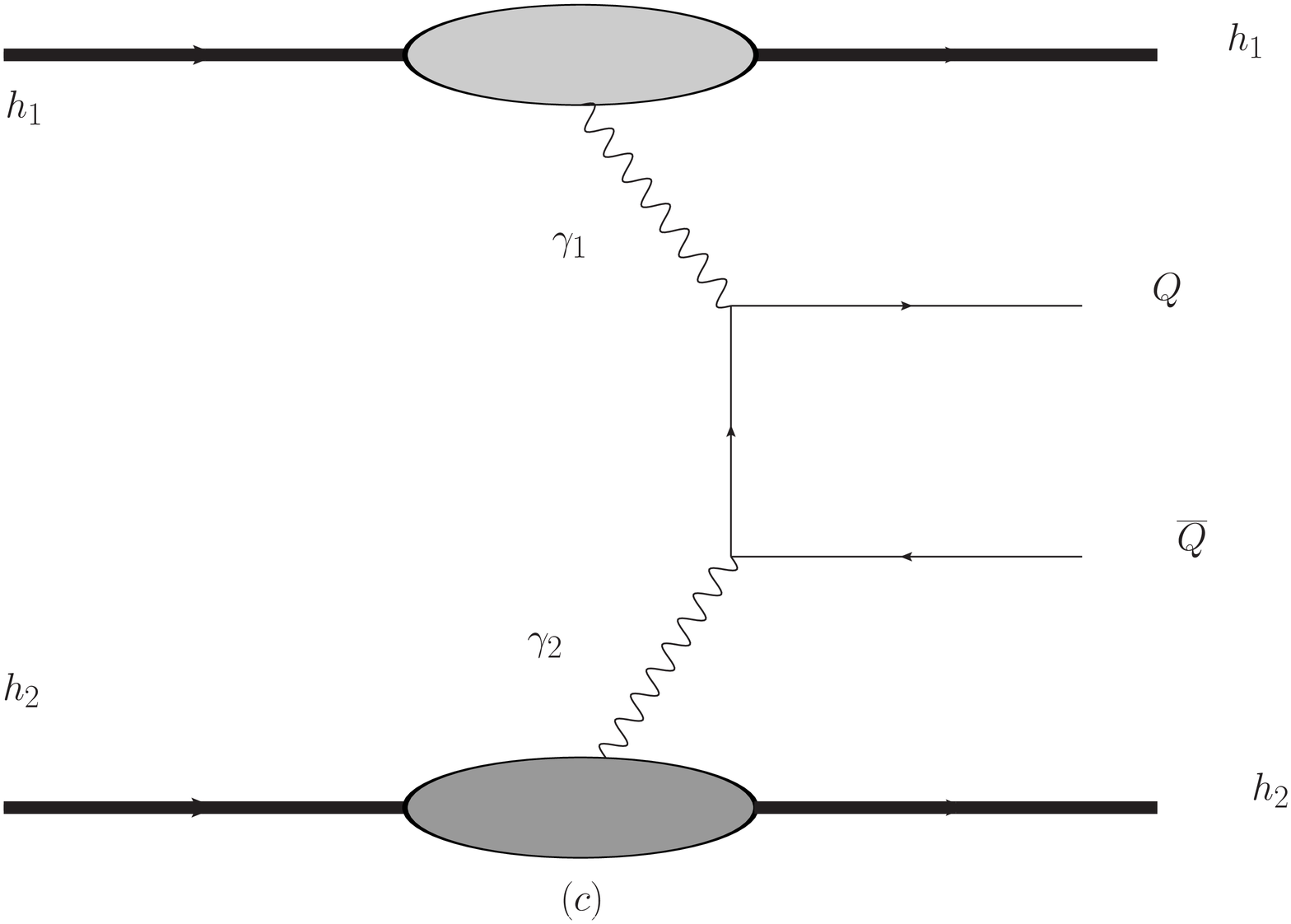}}\\
& & \\
\scalebox{0.18}{\includegraphics{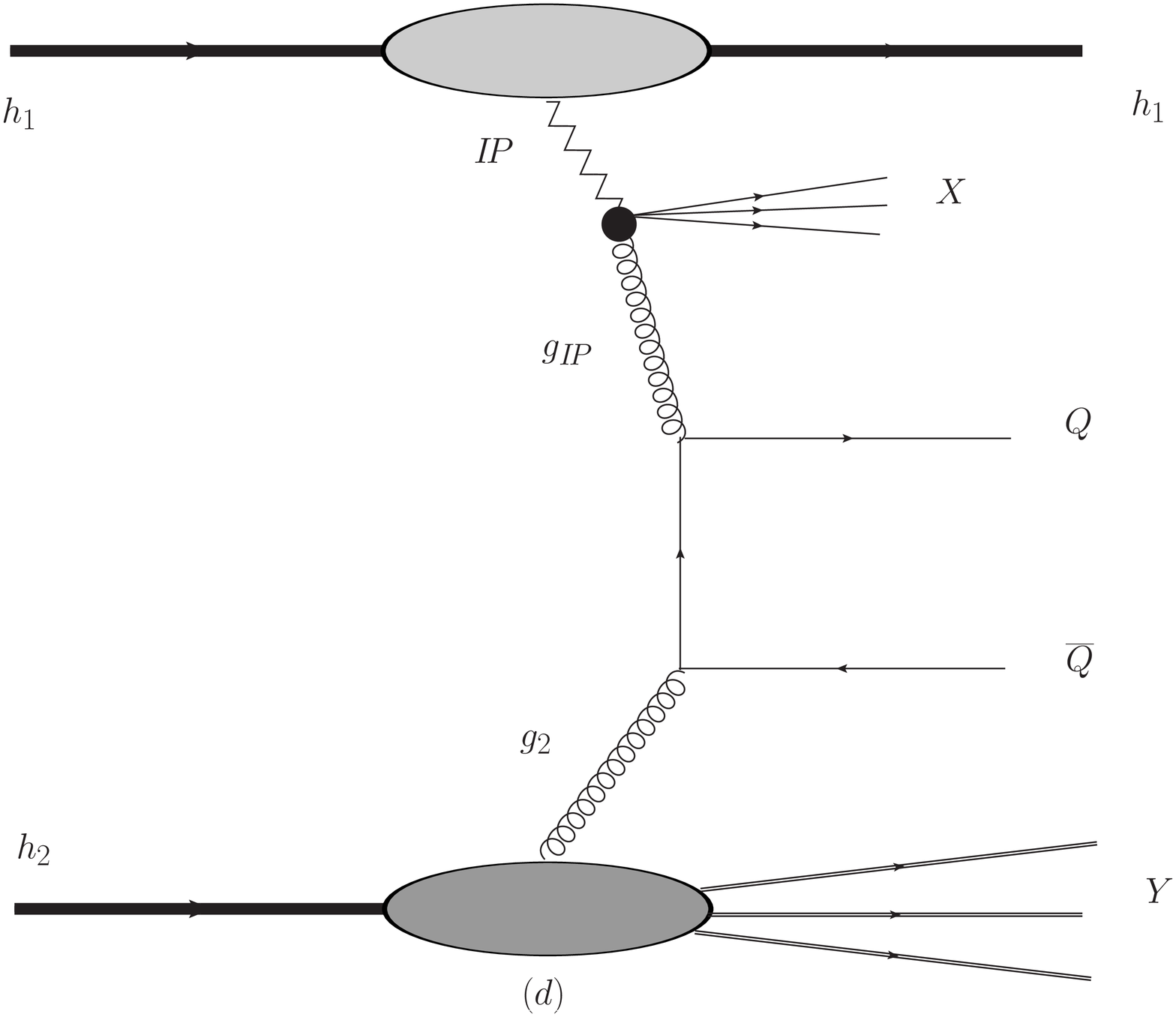}}&\scalebox{0.18}{\includegraphics{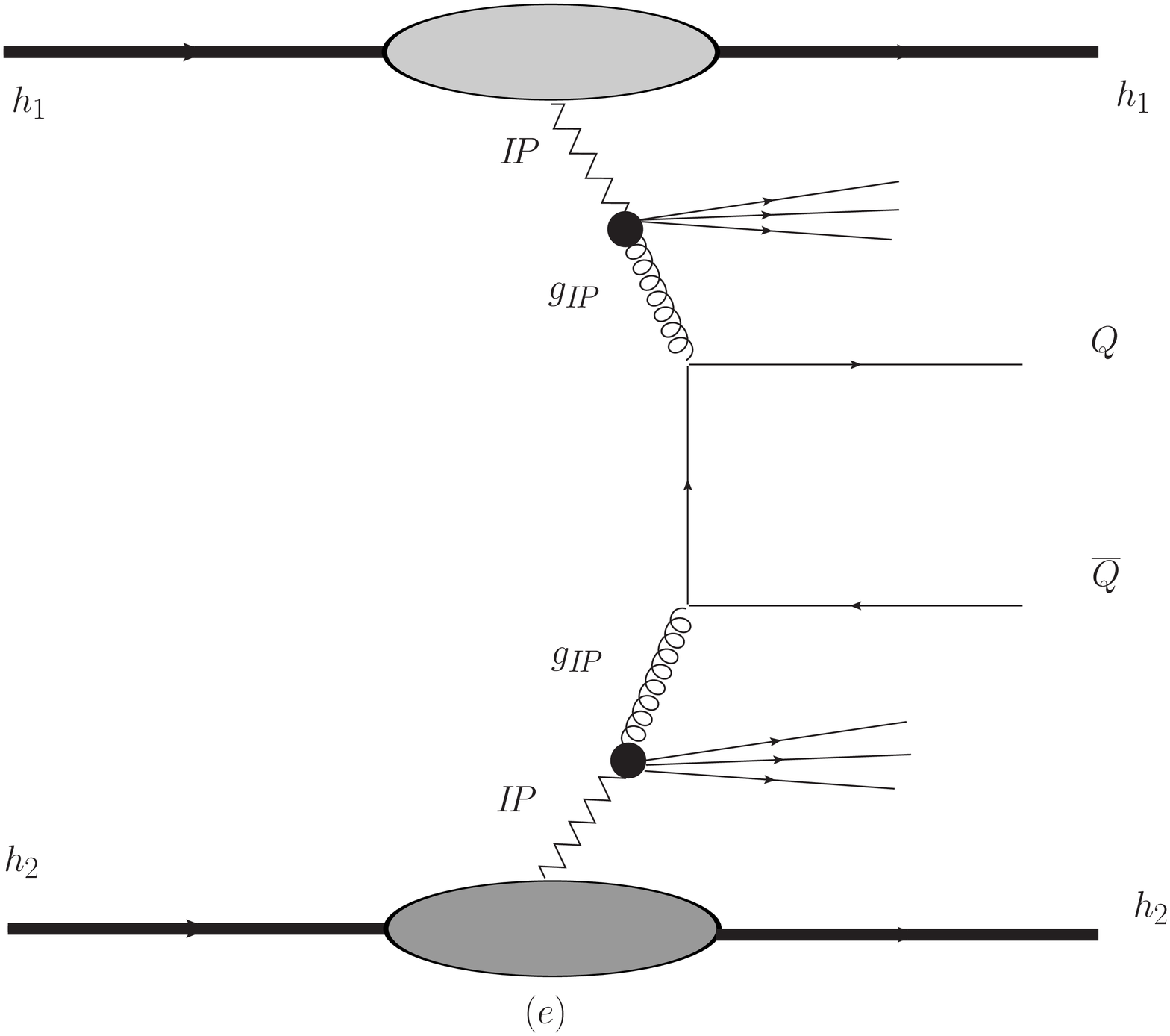}} &\scalebox{0.18}{\includegraphics{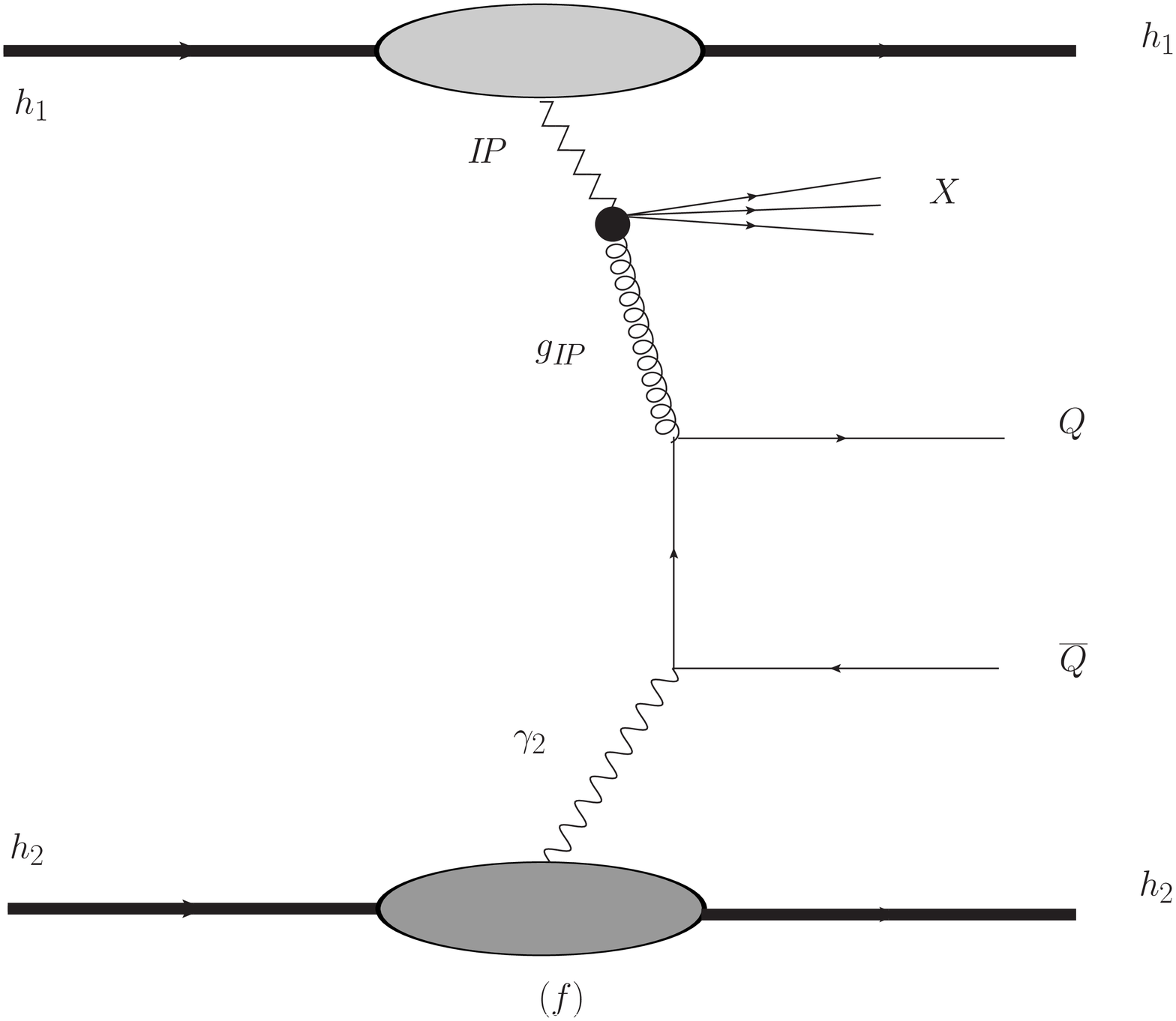}}
\end{tabular}
\caption{Heavy quark production in (a) gluon -- gluon, (b) photon -- gluon, (c) photon -- photon, (d) pomeron -- gluon, (e) pomeron -- pomeron and (f) pomeron -- photon interactions in hadronic collisions.}
\label{Fig:dia}
\end{center}
\end{figure}

At high energies, the heavy quark production in hadronic collisions is dominated by gluon -- gluon interactions, represented in Fig.  \ref{Fig:dia} (a), with the dissociation of the incident hadrons. However, a heavy quark pair also can be generated in photon -- gluon [Fig.  \ref{Fig:dia} (b)], 
photon -- photon [Fig.  \ref{Fig:dia} (c)], pomeron -- gluon [Fig.  \ref{Fig:dia} (d)], pomeron -- pomeron [Fig.  \ref{Fig:dia} (e)] and pomeron -- photon [Fig.  \ref{Fig:dia} (f)] interactions.
These processes can be classified by the topology of the final state, with the presence of one or two 
empty regions in pseudo-rapidity, called rapidity gaps, separating the intact very forward hadron from the central massive object. Taking into account that the photon and the pomeron are color singlet objects, we have that the processes (b) and (d) are characterized by one rapidity gap in the final state, while the processes (c), (e) and (f) are characterized by two. Moreover, the process (c) is a typical example of a exclusive process, where  nothing else is produced except the leading hadrons and the central object. In contrast, if we assume that the Pomeron has a partonic structure, we have that in the processes (e) and (f),  soft particles,   associated to the remnants of the Pomeron, are created accompanying the production of a hard diffractive object. The photon and pomeron -- induced processes are expected to generate emerging protons with different transverse momentum distributions, with those associated to pomeron -- induced having larger transverse momentum. Consequently, in principle it is possible to introduce a selection criteria  to separate these two processes.
 In the last years, these different processes have been studied separately by several authors \cite{Engel,MMM1,antoni,antoni2,victor}, considering different approximations and assumptions, which difficult the direct comparison between its predictions. In particular, the feasibility of the experimental separation between the different contributions for the heavy quark production in photon and pomeron induced interactions  at the LHC still is an open question, mainly during the Run II due to the large pile up expected at large luminosities.

In order to obtain realistic predictions for the heavy quark production in photon and pomeron induced interactions   and to be able to include in the calculations the experimental cuts, the treatment of these processes in a Monte Carlo simulation is  fundamental. Some years ago, the  Forward Physics Monte Carlo (FPMC) was proposed \cite{fpmc}  in order to simulate the central particle production with one or two leading protons and some hard scale in the event. 
In its original version, the pomeron -- gluon, pomeron -- pomeron and photon -- photon interactions in hadronic collisions were implemented considering the  elementary partonic processes as given in HERWIG. Recently, we have generalized this Monte Carlo in order to include photon -- gluon and photon -- pomeron interactions, which allows to estimate the contribution of the different processes presented in Fig. \ref{Fig:dia} in a common framework. In this paper we will present our results for the bottom production at the LHC. In particular, we will perform a comprehensive analysis of the transverse momentum and pseudo -- rapidity distributions for the different processes. It is important to emphasize that similar studies can be performed using the FPMC for other hard processes at the LHC and/or future colliders.

The content of this paper is organized as follows. In the next section we present a brief review of the formalism for the heavy quark production in photon and pomeron -- induced interactions in hadronic collisions. In Section \ref{results} we present our predictions for the pseudo -- rapidity and transverse momentum distributions as well as for the total cross sections for the bottom production in inclusive $pp$ collisions 
and $p\pom$, $\pom \pom$, $\gamma p$, $\gamma \pom$ and $\gamma \gamma$ interactions. Finally, in Section \ref{conc} we summarize our main conclusions.

\section{Heavy quark production in photon and pomeron -- induced interactions}
\label{sec:hq}

The heavy quark production at high energies in inclusive hadronic collisions, represented in Fig. \ref{Fig:dia} (a), can be described at leading order in the collinear factorization formalism by the following expression
\begin{eqnarray}
\sigma(h_1 h_2 \rightarrow X Q\bar{Q} Y) = \int dx_1 \int dx_2 \, g_1(x_1,\mu^2) \cdot g_2(x_2,\mu^2) \cdot \hat{\sigma}(g g \rightarrow Q \bar{Q}) \,\,,
\label{inc}
\end{eqnarray} 
where $x_i$ are the momentum fractions carried by the particles emitted by the incident hadrons, $g_i$ are the gluon density in the incoming hadrons, $\mu$ is the factorization scale  and 
$\hat{\sigma}$ is the partonic cross section for the subprocess  $g g \rightarrow Q \bar{Q}$, calculable using perturbative QCD, which is dominant at high energies. However, 
 a ultra relativistic charged hadron (proton or nuclei)
 give rise to strong electromagnetic fields, such that the photon stemming from the electromagnetic field
of one of the two colliding hadrons can interact with one photon of
the other hadron (photon - photon process) or can interact directly with the other hadron (photon - hadron
process) \cite{upc,epa}. In these processes the total cross section  can be factorized in
terms of the equivalent flux of photons into the hadron projectiles and the photon-photon or photon-target production cross section.  In particular, the heavy quark production in photon -- gluon  and photon -- photon interactions, represented in Figs. \ref{Fig:dia} (b) and (c),  are described by
\begin{eqnarray}
\sigma(h_1 h_2 \rightarrow h_i \otimes Q\bar{Q} Y) = \int dx_1 \int dx_2 \, [\gamma_1(x_1,\mu^2) \cdot g_2(x_2,\mu^2) + g_1(x_1,\mu^2) \cdot \gamma_2(x_2,\mu^2)] \cdot \hat{\sigma}(\gamma g \rightarrow Q \bar{Q}) 
\label{fotglu}
\end{eqnarray}
and
\begin{eqnarray}
\sigma(h_1 h_2 \rightarrow h_1 \otimes Q\bar{Q} \otimes h_2) = \int dx_1 \int dx_2 \, \gamma_1(x_1,\mu^2) \cdot \gamma_2(x_2,\mu^2) \cdot \hat{\sigma}(\gamma \gamma \rightarrow Q \bar{Q}) \,\,,
\label{fotfot}
\end{eqnarray}
respectively. Here $\otimes$ represents the presence of a rapidity gap in the final state and $h_i$ in Eq. (\ref{fotglu}) represents the hadron that have emitted the photon. The basic ingredient in the analysis of these photon - induced processes is the description of the  equivalent photon distribution of the hadron, given by $\gamma(x,\mu^2)$, where $x$ is the fraction of the hadron energy carried by the photon and $\mu$ has to be identified with a momentum scale of the  process. The equivalent photon approximation of a charged  pointlike fermion was formulated  many years ago by Fermi \cite{Fermi} and developed by Williams \cite{Williams} and Weizsacker \cite{Weizsacker}.  In contrast, the calculation of the photon distribution of the hadrons still is a subject of debate, due to the fact that they are not pointlike particles. In this case it is necessary to distinguish between the  elastic and inelastic components.  The elastic component, $\gamma_{el}$, can be estimated analysing the transition $h \rightarrow \gamma h$ taking into account the effects of the hadronic form factors, with the hadron remaining intact in the final state \cite{epa,kniehl}. In contrast, the inelastic contribution, $\gamma_{inel}$, is associated to the transition $h \rightarrow \gamma X$, with $X \neq h$, and  can be estimated taking into account the partonic structure of the hadrons, which can be a source of photons. In what follows we will consider the contribution associated to elastic processes, where the incident hadron remains intact after the photon emission (For a recent discussion about this subject see Refs. \cite{vicgus1,vicgus2}).
A detailed derivation of the  elastic photon distribution of a nucleon was presented in Ref. \cite{kniehl} which can be written as
\begin{eqnarray}
\gamma_{el} (x) = - \frac{\alpha}{2\pi} \int_{-\infty}^{-\frac{m^2x^2}{1-x}} \frac{dt}{t}\left\{\left[2\left(\frac{1}{x}-1\right) + \frac{2m^2x}{t}\right]H_1(t) + xG_M^2(t)\right\}\,\,,
\label{elastic}
\end{eqnarray}
 where $t = q^2$ is the momentum transfer squared of the photon, 
 \begin{eqnarray}
 H_1(t) \equiv \frac{G_E^2(t) + \tau G_M^2(t) }{1 + \tau}
 \end{eqnarray}
 with $\tau \equiv -t/m^2$, $m$ being the nucleon mass, and where $G_E$ and $G_M$ are the Sachs elastic form factors.
 Although an analytical expression for the elastic component is presented in Ref. \cite{kniehl}, it is common to found in the literature the study of photon - induced processes considering an approximated expression proposed in Ref. \cite{dz}, which can be obtained from Eq.~(\ref{elastic}) by disregarding the contribution of the magnetic dipole moment and the corresponding magnetic form factor. As demonstrated in Ref. \cite{vicwerdaniel} the difference between the full and the approximated expression is smaller than 5\% at low-$x$. Consequently, in what follows we will use the expression proposed in Ref. \cite{dz}, where the elastic photon distribution is given by
 \begin{eqnarray}
\gamma_{el}(x)&=&\frac{\alpha}{\pi}\left(\frac{1-x+0.5x^{2}}{x}\right) \times  \left[\ln(\Omega)
-\frac{11}{6}+\frac{3}{\Omega}-\frac{3}{2\Omega^{2}}+\frac{1}{3\Omega^{3}}\right]\,\,,
\label{dz}
\end{eqnarray}
where $\Omega=1+(0.71 \mathrm{GeV}^{2})/Q_{min}^{2}$ and 
 $Q^2_{min} \approx (x m)^2/(1-x)$.

Assuming that the diffractive processes, represented in Figs. \ref{Fig:dia} (d) -- (f), can be described the resolved Pomeron model, we have that 
cross sections can be written similarly to the inclusive heavy quark production, with a diffractive gluon distribution $g^D (x,\mu^2)$ replacing the standard inclusive gluon distribution (See e.g. Refs. \cite{martinwus,vicmag_cha}). Moreover, it is assumed that the diffractive cross sections can be expressed in terms of parton distributions in the Pomeron and a Regge parametrization of the flux factor describing the Pomeron emission. The  parton distributions have evolution given by the DGLAP evolution equations and   are determined from events with a rapidity gap or a intact proton, mainly at HERA. 
Explicitly one have that the heavy quark production in pomeron -- gluon (single diffractive) and pomeron -- pomeron (double diffractive) interactions will be described by
\begin{eqnarray}
\sigma(h_1 h_2 \rightarrow h_i \otimes X Q\bar{Q} Y) = \int dx_1 \int dx_2 \, [g^D_1(x_1,\mu^2) \cdot g_2(x_2,\mu^2) + g_1(x_1,\mu^2) \cdot g^D_2(x_2,\mu^2)] \cdot \hat{\sigma}(g g \rightarrow Q \bar{Q}) 
\label{pomglu}
\end{eqnarray}
and
\begin{eqnarray}
\sigma(h_1 h_2 \rightarrow h_1 \otimes X Q\bar{Q} Y \otimes h_2) = \int dx_1 \int dx_2 \, g^D_1(x_1,\mu^2) \cdot g^D_2(x_2,\mu^2) \cdot \hat{\sigma}(g g \rightarrow Q \bar{Q}) \,\,,
\label{pompom}
\end{eqnarray}
respectively, with $h_i$ in Eq. (\ref{pomglu}) representing the hadron that have emitted the Pomeron.
Similarly, the cross section for the heavy quark production in pomeron -- photon interactions 
is given by
\begin{eqnarray}
\sigma(h_1 h_2 \rightarrow h_1 \otimes Q\bar{Q} X \otimes h_2 ) = \int dx_1 \int dx_2 \, [g^D_1(x_1,\mu^2) \cdot \gamma_2(x_2,\mu^2) + \gamma_1(x_1,\mu^2) \cdot g^D_2(x_2,\mu^2)] \cdot \hat{\sigma}(\gamma g \rightarrow Q \bar{Q}) \,\,. 
\label{pompho}
\end{eqnarray}
In the resolved Pomeron model, 
the diffractive gluon distribution in the proton, $g^D (x,\mu^2)$, is defined as a convolution of the Pomeron flux emitted by the proton, $f_{I\!\!P}(x_{I\!\!P})$, and the gluon distribution in the Pomeron, $g_{I\!\!P}(\beta, \mu^2)$,  where $\beta$ is the momentum fraction carried by the partons inside the Pomeron. 
The Pomeron flux is given by $f_{I\!\!P}(x_{I\!\!P})= \int_{t_{min}}^{t_{max}} dt f_{\pom/p}(x_{{I\!\!P}}, t)$, where $f_{\pom/p}(x_{\pom}, t) = A_{\pom} \cdot \frac{e^{B_{\pom} t}}{x_{\pom}^{2\alpha_{\pom} (t)-1}}$ and $t_{min}$, $t_{max}$ are kinematic boundaries. The Pomeron flux factor is motivated by Regge theory, where the Pomeron trajectory assumed to be linear, $\alpha_{\pom} (t)= \alpha_{\pom} (0) + \alpha_{\pom}^\prime t$, and the parameters $B_{\pom}$, $\alpha_{\pom}^\prime$ and their uncertainties are obtained from fits to H1 data  \cite{H1diff}. 
The diffractive gluon distribution is then given by
\begin{eqnarray}
{ g^D(x,\mu^2)}=\int dx_{I\!\!P}d\beta \delta (x-x_{I\!\!P}\beta)f_{I\!\!P}(x_{I\!\!P})g_{I\!\!P}(\beta, \mu^2)={ \int_x^1 \frac{dx_{I\!\!P}}{x_{I\!\!P}} f_{I\!\!P}(x_{I\!\!P}) g_{I\!\!P}\left(\frac{x}{x_{I\!\!P}}, \mu^2\right)}
\end{eqnarray}
Similar definition can be established for the diffractive quark distributions. However, in what follows we will disregard the quark contributions for the heavy quark production, since they are negligible at high energies.
In our analysis we use the diffractive gluon distribution obtained by the H1 Collaboration at DESY-HERA, denoted fit A in Ref. \cite{H1diff}. Moreover, we use the inclusive gluon distribution as given by the CT10 parametrization \cite{cteq10}.

In order to obtain reliable predictions for the single and double diffractive cross sections, associated to pomeron -- gluon and pomeron -- pomeron interactions,  one should take into account that the QCD hard scattering factorization theorem for diffraction is violated in $pp$ collisions by soft interactions which lead to an extra production of particles that destroy the rapidity gaps related to pomeron exchange. The inclusion of these additional absorption effects can be parametrized in terms of a rapidity gap survival probability, $S^2$, which corresponds to the probability of the scattered proton not to dissociate due to the secondary interactions. These effects have been calculated considering different approaches giving distinct predictions (See, e.g. Ref. \cite{review_martin}). An usual approach in the literature is the calculation of an average probability  
$\langle |S|^2\rangle$ and after to multiply  the cross section by this value. As previous studies for single and double diffractive production \cite{MMM1,antoni,antoni2,cristiano1,cristiano2} we also follow this simplified approach assuming $\langle |S|^2\rangle = 0.05$ for single diffractive processes and  $\langle |S|^2\rangle = 0.02$ for double diffractive processes. These values are taken from Ref. \cite{KMR}. In contrast, for photon -- induced interactions we will assume  $\langle |S|^2\rangle = 1$. However, it is important to emphasize that the magnitude of the rapidity gap survival probability in $\gamma \pom$ still is an open question. For example, in Ref. \cite{Schafer}  the authors have estimated $\langle |S|^2\rangle$ for the exclusive photoproduction of $J/\Psi$ in  $pp/p\bar{p}$ collisions, obtaining that it is $ \sim 0.8 - 0.9$ and depends  on the  rapidity of the vector meson (See also Refs. \cite{Guzey,Martin}).

\begin{figure}[t]
\begin{center}
\scalebox{0.45}{\includegraphics{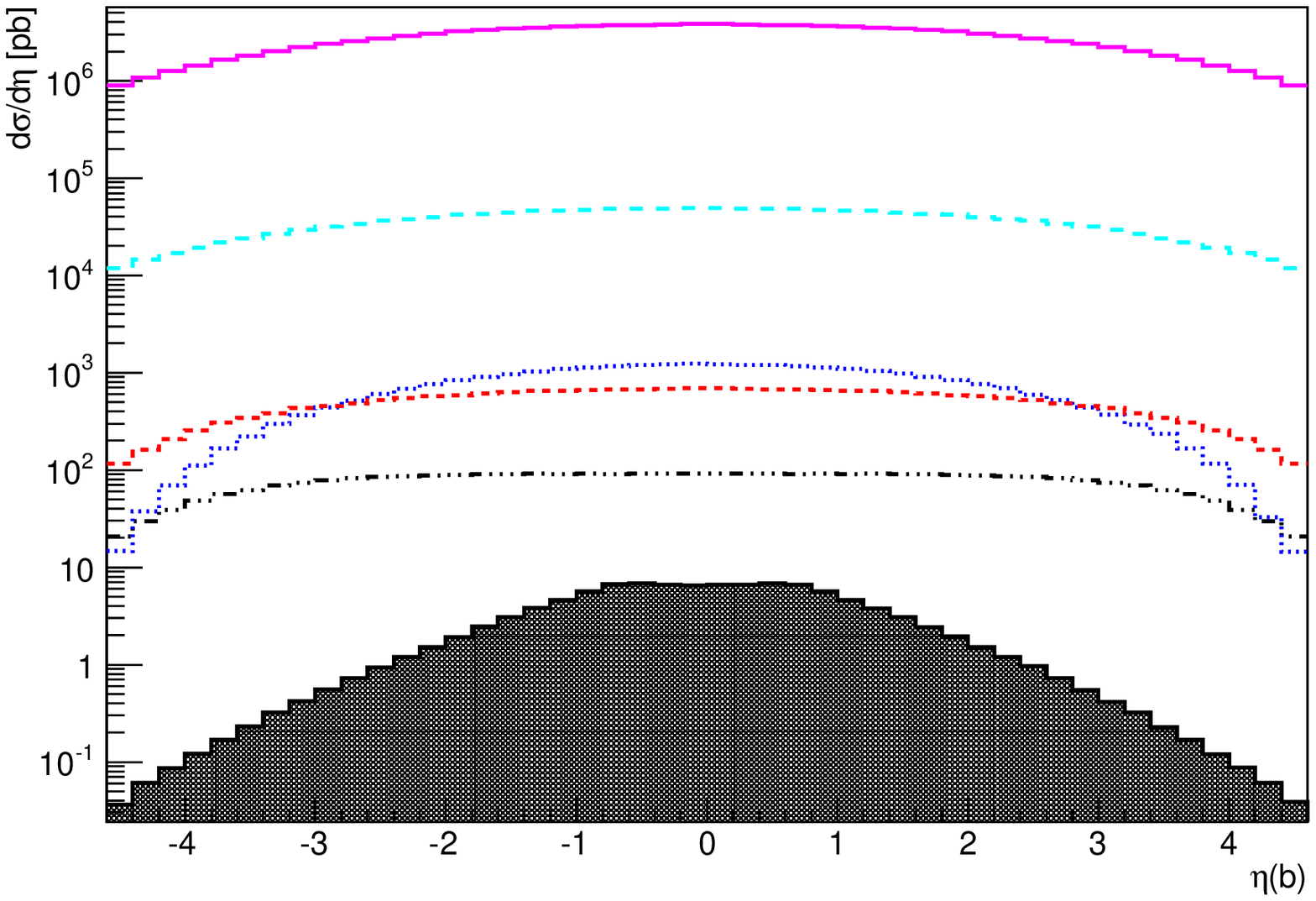}}
\scalebox{0.45}{\includegraphics{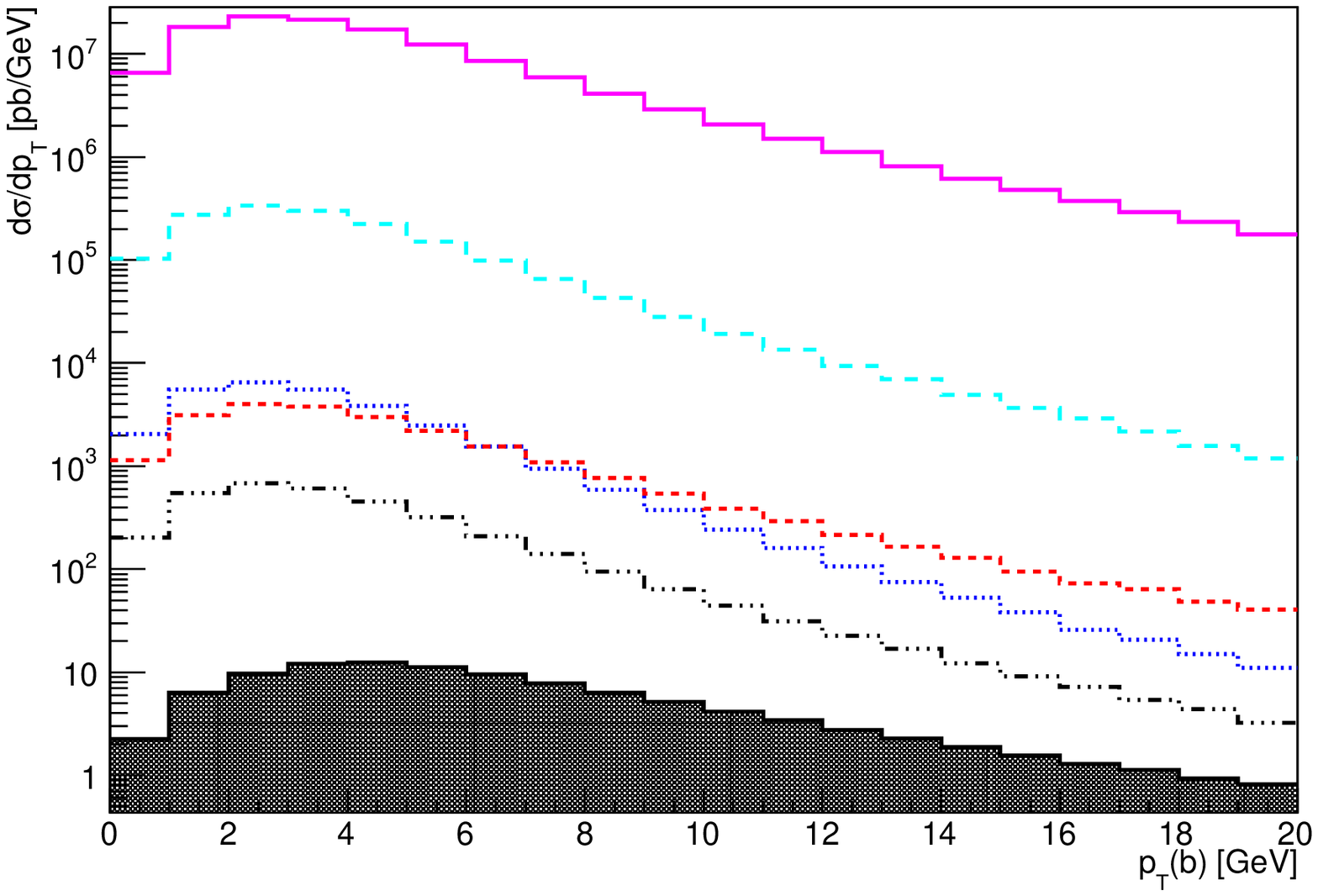}}
\scalebox{0.45}{\includegraphics{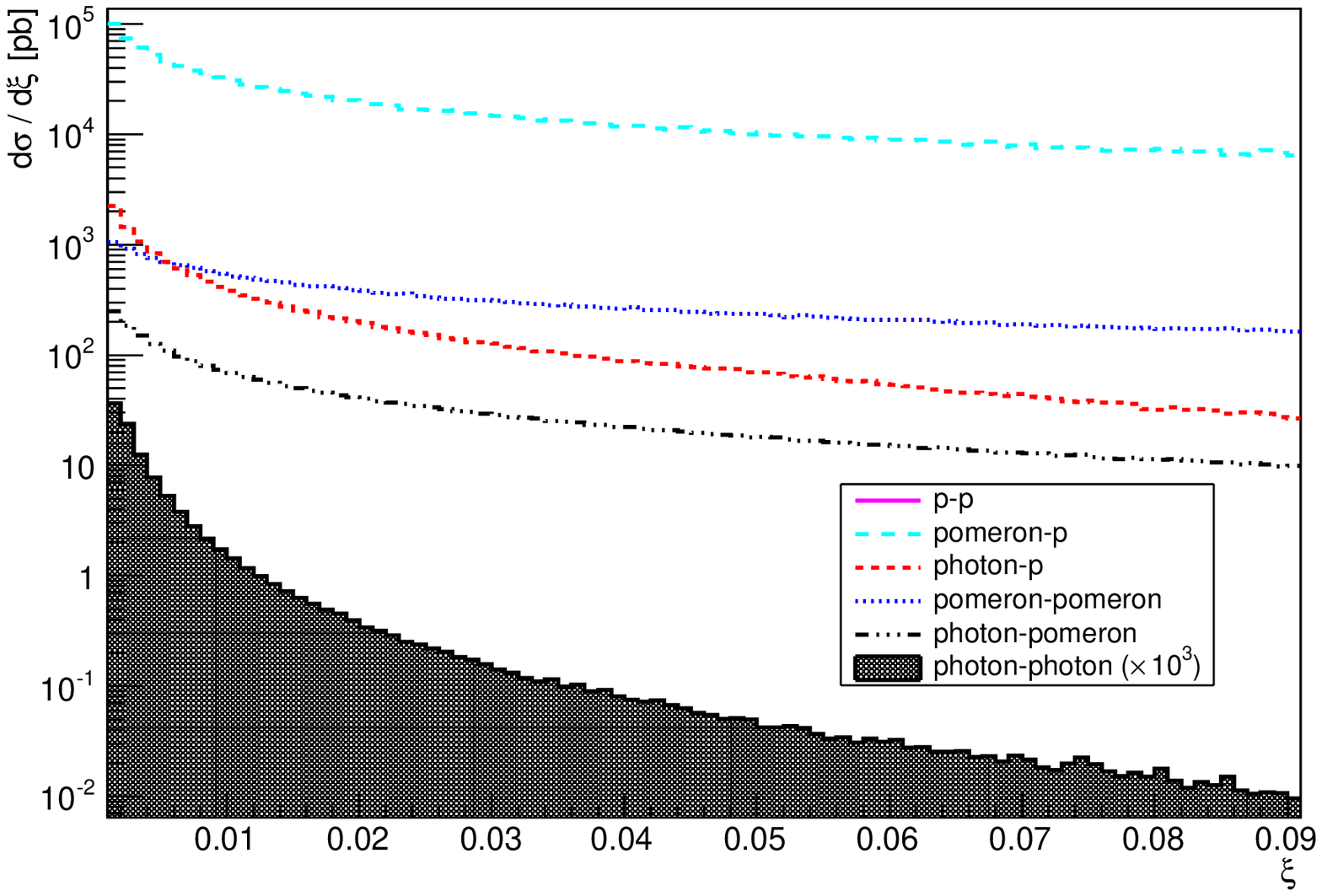}}
\caption{Predictions for the pseudo - rapidity (upper panel), transverse momentum (middle panel) and $\xi$ (lower panel) distributions for the bottom production in inclusive $pp$ collisions and $\pom p$, $\pom \pom$, $\gamma p$, $\gamma \pom$ and $\gamma \gamma$ interactions.}
\label{Fig:lhc}
\end{center}
\end{figure}

\begin{figure}[t]
\begin{center}
\scalebox{0.45}{\includegraphics{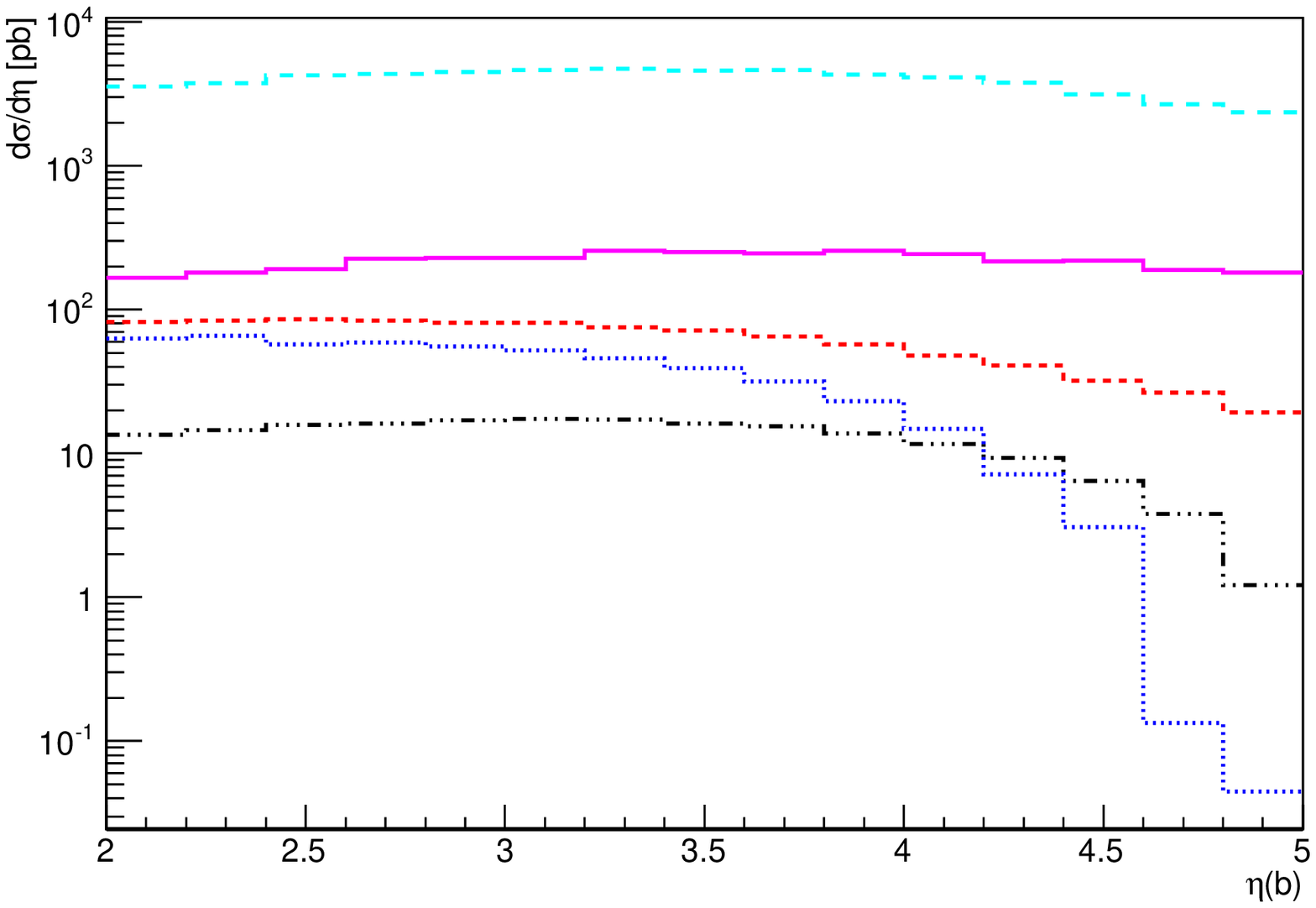}}
\scalebox{0.45}{\includegraphics{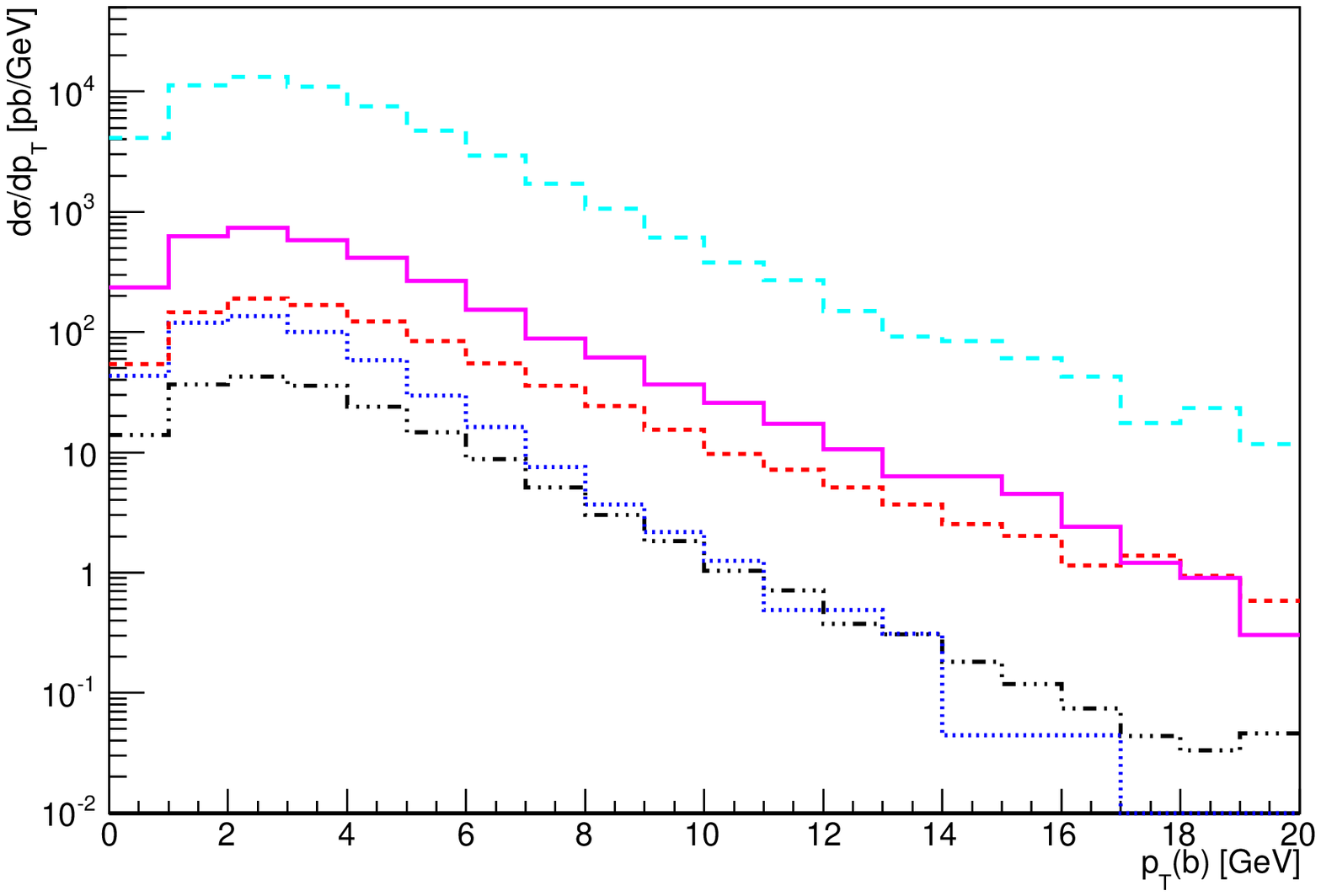}}
\scalebox{0.45}{\includegraphics{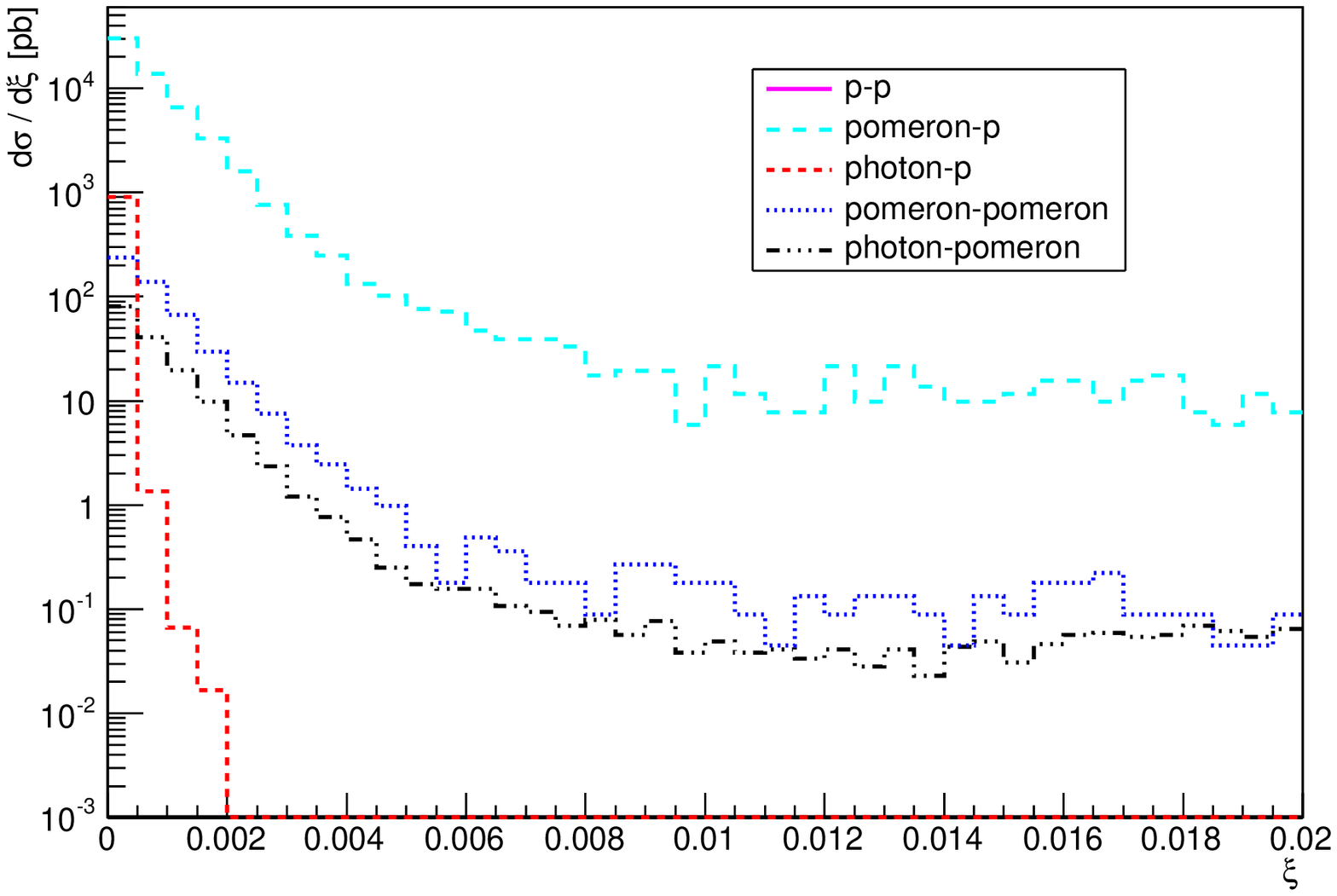}}
\caption{Predictions for the pseudo - rapidity (upper panel), transverse momentum (middle panel) and $\xi$ (lower panel) distributions for the bottom production in inclusive $pp$ collisions and $\pom p$, $\pom \pom$, $\gamma p$, $\gamma \pom$ and $\gamma \gamma$ interactions considering the presence of a rapidity gap in the detector acceptance of the LHCb experiment.}
\label{Fig:lhcb}
\end{center}
\end{figure}

\section{Results}
\label{results}

In what follows we present our results for the bottom  production at LHC energy (For a similar analysis for the charm production see  Ref. \cite{antoni}). We assume $\sqrt{s} = 8$ TeV and $m_b = 4.5$ GeV. The cross sections for the partonic subprocesses are calculated at leading order in FPMC using HERWIG 6.5. In Fig. \ref{Fig:lhc} we show our results for the pseudo -- rapidity $\eta(b)$ (upper panel) and  transverse momentum $p_T (b)$ (middle panel) 
distributions, presenting separately the predictions for the bottom production in inclusive $pp$ interactions as well as in photon and pomeron -- induced interactions. For 
$p\pom$, $\pom \pom$, $\gamma
 p$, $\gamma \pom$ and $\gamma \gamma$ interactions   
 we also present the $\xi$ distributions in the lower panel.
Assuming that the different events can be separated by requiring the presence of one or two rapidity gaps in the final state, we have that the events with one rapidity gap are dominated by single diffractive $p\pom$ interactions, with the contribution of the $\gamma p$ being smaller by factor $10^2$. For the events with two rapidity gaps in the final state, the bottom production is dominated by double diffractive $\pom \pom$ interactions, with the contributions of the $\gamma \pom$ and $\gamma \gamma$ processes being smaller by a factor $\approx 10$     and $\approx 10^6$, respectively. 
Consequently, if the rapidity gaps could be detected, the analysis of the bottom quark production is useful for the study of the single and double diffractive processes. However, due to the non - negligible pileup present at the LHC, it is not an easy task. Another possibility to separate photon and pomeron -- induced processes is the detection of the outgoing intact protons. Recently, the 
ATLAS, CMS and TOTEM Collaborations have proposed the setup of forward detectors \cite{Albrow:2008pn,ctpps,marek}, which will enhance the kinematic coverage for such investigations. However, in a first moment only one of the very forward detectors  will be installed, which would not eliminate the one rapidity events associated to $\gamma p$ and $p\pom$ interactions, where one of the protons dissociates.  
In this case the bottom production in single diffractive events is dominant, with the production in $\gamma p$ interactions being similar to that in $\pom \pom$ one. In particular, our results demonstrate that the bottom production at large rapidities and large transverse momentum is larger in 
$\gamma p$ than in $\pom \pom$ interactions. Moreover, the $\gamma \pom$ and $\pom \pom$ processes predict similar number of events at large rapidities.  Regarding the $\xi$ distributions, we obtain that the events are dominated by single diffractive processes, with the $\pom \pom$ one being the second more important process at large $\xi$. On the other hand, at small $\xi$ the $\gamma p$ process contributes more than the $\pom \pom$ one. Finally, we obtain that the bottom production by $\gamma \gamma$ interactions is negligible in comparison to the other photon and pomeron induced processes.

The previous results indicate that the contribution of the pomeron - induced interactions for bottom production are very high. In what follows we analyse the possibility to study these processes with the Run I data by the LHCb Collaboration. Our main motivation for this analysis is associated to the fact that the LHCb detector \cite{lhcb_det} is fully instrumented in the forward acceptance designed for the study of particles containing the bottom or charm quarks. The experiment  is also able to reject activity in the backward region using tracks reconstructed in the Vertex Locator (VELO) sub-detector. The experiment runs at lower instantaneous luminosity benefiting from low pile-up conditions. Indeed, LHCb has already published central exclusive analyses exploring the ability of requiring a backward gap \cite{lhcb,lhcb2,lhcb_ups}. Since the cross-sections for diffractive bottom production are very high, we propose an analysis with Run I data to study in detail the single diffraction (SD) component. To prove that one can obtain a pure sample of bottom SD production, we  will select bottom quarks within the LHCb acceptance $2.0<\eta<5.0$ and require no charged particles in the backward region of $-4.5<\eta<-1.5$.
The predictions for the pseudo - rapidity, transverse momentum and $\xi$ distributions are presented in Fig. \ref{Fig:lhcb}. As the contribution associated to $\gamma \gamma$ interactions is very small, it is not included in the figures. Our results indicate that the single diffractive contribution becomes dominant, with the background associated to inclusive bottom production  being smaller by a factor 20. Comparing  the photon and pomeron - induced processes, we have that the main subleading contribution is associated to $\gamma p$ interactions, which dominates over the $\pom \pom$ and $\gamma \pom$ contributions at large pseudo - rapidities and transverse momentum as well as at small $\xi$. Moreover, our results indicate that the $\pom \pom$ and $\gamma \pom$ contributions are similar.

\begin{table}[t]
\begin{center}
 \vspace{0.5cm}
\begin{tabular}{|c|c|c|c|c|c|c|}
\hline
\hline
 & $pp$ & $p\pom$ & $\pom\pom$ & $\gamma p$ & $\gamma\pom$ & $\gamma\gamma$  \\
\hline
\hline
LHC & 3.59 $\times 10^{8}$  & 2.63 $\times 10^{7}$ & 1.51 $\times 10^{6}$ & 11474.80 & 1744.91 & 0.11  \\
\hline
LHCb & $1.53 \times 10^7$ &  210036.10 & 1846.33 & 3065.27 & 573.14 & 7.97 $\times 10^{-7}$  \\
\hline
LHCb gap & 3725.05 &  59392.50 & 518.22 & 932.01 & 189.37 & 7.97 $\times 10^{-7}$  \\
  \hline
\hline
\end{tabular}
\caption{Total cross sections in $pb$ for the bottom production in inclusive $pp$ collisions and 
and $p\pom$, $\pom \pom$, $\gamma p$, $\gamma \pom$ and $\gamma \gamma$ interactions. 
The LHCb gap line represents the results obtained considering the detector acceptance with a rapidity gap requirement in the LHCb experiment.}
\label{tab1}
\end{center}
\end{table}

In Table~\ref{tab1} we present our predictions for the total cross sections considering the inclusive $pp$ collisions and the photon and pomeron - induced interactions. If we consider the full LHC kinematical range, denoted LHC in the Table, one observes a reduction of one order of magnitude when going from the inclusive to single diffractive, and from single to double diffractive. However, as verified in Table~\ref{tab1}, the photon and pomeron induced cross sections are still sizeable and could be measured at the LHC. A similar conclusion is valid by the analysis of our predictions for the bottom production in the the kinematical range of the LHCb detector, denoted by LHCb in the Table.
Finally, in the line denoted LHC gap, we present our predictions for the cross sections when we require the presence of a rapidity gap in the detector acceptance of the LHCb detector, as discussed in the previous paragraph. In this case, after selection of diffractive events in the LHCb detector, we find a strong suppression of the inclusive processes and the single diffractive events become dominant. Basically, we obtain that  a small number of inclusive events are produced with a rapidity gap in the kinematical range of the LHCb detector, which are a background for genuine diffractive events.  Assuming the trigger efficiency for B-hadrons at LHCb is $40\%$ and the fraction of data with only one primary vertex is of about $20\%$~\cite{lhcbperformace}, the number of $p\pom$ events in Run I ($2\mbox{fb}^{-1}$) is predicted to be $9.5\times10^{6}$ with almost no background. For higher $p_{T}$ bottom quarks, b-jets may also be used with similar efficiencies~\cite{lhcbjet}. This result suggests that studies of this process using the Run I data is feasible for several B-hadron decay channels and high-$p_T$ b-jets.

\section{Summary}
\label{conc}

As a summary, in this paper we have presented a detailed analysis  for the bottom production  in $pp$ collisions at the LHC. In particular, the comparison between the predictions for the photon and pomeron - induced interactions was presented  considering a common framework implemented in the Forward Physics Monte Carlo. We have generalized this Monte Carlo for photon - induced processes and performed a detailed comparison between the 
$p\pom$, $\pom \pom$, $\gamma p$, $\gamma \pom$ and $\gamma \gamma$ predictions for the bottom production in $pp$ collisions at $\sqrt{s} = 8$ TeV. For the pomeron - induced processes, we have considered the  framework of the Resolved Pomeron model corrected for absorption effects, as used in the estimation of several other diffractive processes.  Our results indicate that the single and double diffractive processes are dominant if the outgoing protons can be tagged in the final state. On the other hand, if only one proton is tagged, the contribution of the photon - induced processes cannot be disregarded. Finally, we have analysed the possibility of study of the diffractive processes  considering the experimental data obtained in the Run I and demonstrated that this study is feasible using the detector acceptance of the LHCb detector. Such experimental analysis  would help to constrain the underlying model for the Pomeron and the absorption corrections, which are important open questions in Particle Physics.

\begin{acknowledgments}
Useful discussions with Christophe Royon are gratefully acknowledged. This research was supported by CNPq, CAPES and FAPERGS, Brazil. 
\end{acknowledgments}

\end{document}